\documentclass[aps, showpacs,pre,unsortedaddress,eqsecnum,twocolumn] {revtex4-1}
\usepackage{hyperref}
\usepackage{amsmath }
\usepackage{amssymb}
\usepackage{epsfig}
\usepackage{natbib}
\newcommand*{\be}{\begin{equation}}
\newcommand*{\ee}{\end{equation}}
\newcommand*{\bea}{\begin{eqnarray}}
\newcommand*{\eea}{\end{eqnarray}}

\begin{document}

\title[]{Stability of cnoidal waves in the parametrically driven nonlinear Schr\"odinger equation}

\author{I. V. Barashenkov}
\email{igor.barashenkov@gmail.com} 

\author{M. A. Molchan}
\email{m.moltschan@gmail.com}
\affiliation{Department of Mathematics and Applied Mathematics,
University of Cape Town, Private Bag Rondebosch 7701, South
Africa}
\affiliation{National Institute for Theoretical Physics (NITheP),
Stellenbosch, South Africa}

\begin{abstract}
The
parametrically driven, damped nonlinear Schr\"odinger equation has two
cn- and two dn-wave solutions. We show that 
 one pair of the cn and dn  solutions is 
 unstable for any combination  of the driver's strength, dissipation coefficient and
spatial period of the wave; this instability is against periodic perturbations.
The second dn-wave solution is shown to be unstable
against antiperiodic perturbations --- in a certain region of the parameter space.
We also consider   quasiperiodic perturbations with long modulation wavelength,
in the limit where the driving strength is only weakly exceeding the damping coefficient.

\end{abstract}

\maketitle

\section{Introduction}

Periodic waves arise in media of different physical nature. They
find direct applications in plasma physics~\cite{Mamaev,Tikh},
nonlinear optics~\cite{opt1,opt2,opt3}, solid state
physics~\cite{solid1,solid2}, and physics of Bose-Einstein
condensates~\cite{Bose}. 
The  nonlinear periodic waves
(or \emph{cnoidal} waves) interpolate between plane waves 
and solitons.

One of the most important problems associated with periodic waves
is their stability in various nonlinear media. To date, stable
periodic patterns were discovered in physical settings modelled by
self-defocusing
nonlinearities~\cite{stable1,stable2,stable3,stable4,stable5}, and
in Bose-Einstein condensates confined by periodic
potentials~\cite{BoseSt1,BoseSt2}. Families of stable cnoidal
waves supported by focusing nonlinearities were revealed in
quadratic media~\cite{Foc1,Foc2}. All these results pertain to
conservative media where self-supported structures exist due
to the balance between dispersion and nonlinearity. In the presence of
dissipation, an additional balance  between gain and loss
is needed.
 In the dissipative case the parameters of the solution (e.g.,
amplitude and phase) are fixed by the parameters of the governing
equation, whereas in conservative settings solutions form
continuous families. So far, stable stationary dissipative
periodic waves were mostly observed in optical cavities.

In this paper we study stability properties of periodic wave solutions of
the parametrically driven, damped nonlinear Schr\"odinger equation:
\begin{equation}\label{1}
 i \psi_t+\psi_{xx}-\psi+2 \psi|\psi|^2=h\psi^*-i\gamma\psi,\quad h, \gamma >0.
\end{equation}
Here  $\gamma$ is the damping
coefficient and $h$ the amplitude of the parametric driver. 
Equation \eqref{1}
is an archetypal  equation for small and slowly-varying amplitudes of 
waves and patterns in 
spatially-distributed parametrically driven systems.
It arises in a wide variety of physical contexts including 
instabilities in plasma~\cite{plasma1,plasma2}, amplitude
generation in Josephson junctions~\cite{junction1,junction2}, and
signal amplification effects in fiber optics~\cite{ampl1,ampl2}

At the same time, the knowledge about periodic nonlinear waves of
PDNLS is rather scarce. Umeki~\cite{Umeki} examined numerically
stability properties of the families of cn- and dn-waves of PDNLS
considered in the context of water waves in a vertically forced
long container ($h<0$). For a fixed value of the  spatial period
of the wave, he obtained  the corresponding stability diagram and
studied the
 temporal evolution of the perturbation. Umeki's numerical considerations were restricted to
perturbations of the initial wave profile  having the same period
as the cnoidal wave itself.

In this paper, we examine the linear stability of the cn and dn type  solutions of
Eq.~(\ref{1}) with respect to two broad classes of perturbations.
The first class consists of perturbations which are periodic or
antiperiodic with the 
period of the wave. These perturbations have a discrete spectrum
and  are relatively easy to analyse. On the other hand, 
the periodic or antiperiodic perturbations turn out to be sufficient to 
detect the instability of the underlying wave in a wide range of parameter values.

In particular, we prove that out of two cn- and two dn-wave solutions, 
one pair of cn- and dn-waves is
unstable against periodic perturbations --- 
for any combination  of the driver's strength, dissipation coefficient and
spatial period of the wave. 
The other dn wave is shown to be unstable
against antiperiodic perturbations --- in a certain region of the parameter space.

The second class includes quasiperiodic perturbations
which have the form of symmetry eigenvectors modulated 
by a long wavelength. Here, the small wavenumber of the modulation serves as 
a perturbation expansion parameter.
The perturbation theory can be developed near the lower boundary of the 
cnoidal-wave existence domain, that is, for driving strengths only slightly exceeding the damping coefficient. 

Using this perturbation approach, we reproduce several results obtained by other means.
First, we confirm the instability of the cn- and dn-wave solutions of the 
unperturbed ($\gamma=h=0$) NLS equation, a result already available in literature.
Second, we corroborate our own conclusions on the instability of the two damped-driven dn waves 
in the $h \approx \gamma$ case. 
More importantly,  the long-wavelength-modulation treatment provides useful information on the
 structure of the spectrum of the cn wave whose
stability cannot be classified by restricting to the periodic perturbations. 

We complement the perturbation study of the cn wave with a numerical analysis of 
its linearised spectrum in the $h \approx \gamma$ case.

\section{Periodic solutions and their linearizations}
\label{sec_lin}

\subsection{Two pairs of cnoidal waves}
Eq.~(\ref{1}) has two pairs of nonequivalent periodic solutions.
One pair is expressible in terms of the Jacobi cosine function:
\be 
\label{sol1} \Psi_\pm^{\textrm{cn}}=A_\pm q_{\textrm{cn}}
\left( A_\pm x,k \right) \textrm{e}^{ -i\theta_\pm},
\ee 
where
\be
\label{qcn}
q_\textrm{cn}(X,k)=\frac{k}{\sqrt{2k^2-1}}\textrm{cn}\left(\frac{X}{\sqrt{2k^2-1}},k
\right),\ee $1/\sqrt{2}< k\le1$. We will occasionally be referring
to these solutions as $\textrm{cn}^+$ and $\textrm{cn}^-$,
respectively. The other two solutions (referred to as
$\textrm{dn}^+$ and $\textrm{dn}^-$ in what follows) invoke the
Jacobi dn function instead:
\be \label{sol2}\Psi_\pm^\textrm{dn}=A_\pm q_{\textrm{dn}}\left(
A_\pm x,k \right)\textrm{e}^{ -i\theta_\pm}, \ee where
\be\label{qdn}
q_\textrm{dn}(X,k)=\frac{1}{\sqrt{2-k^2}}\textrm{dn}\left(\frac{X}{\sqrt{2-k^2}},k
\right),\ee $ 0\le k\le1$. The amplitudes $A_\pm$ are given, in
both cases, by \be\label{ampl} A_\pm^2=1\pm\sqrt{h^2-\gamma^2},
\ee and the phases by \[
\theta_+=\frac{1}{2}\textrm{arcsin}\left(\frac{\gamma}{h}\right),\
\ \theta_-=\frac{\pi}{2}-\theta_+.
\]

Eq.~(\ref{ampl}) carries the entire information on the domain of
existence of the four periodic solutions on the
($h$,$\gamma$)--parameter plane. For the given $\gamma$, the
cnoidal waves $\Psi_+^{\textrm{cn}}$ and $\Psi_+^{\textrm{dn}}$
exist for all $h>\gamma$ whereas the domain of the solutions
$\Psi_-^{\textrm{cn}}$ and $\Psi_-^{\textrm{dn}}$ is bounded on both sides:
$\gamma<h<\sqrt{1+\gamma^2}$. The two cn solutions have the spatial
periods $ L_\textrm{cn}/A_\pm$,
where 
\be
\label{Lcn} 
L_\textrm{cn}=L_\textrm{cn}(k)=2K(k)\sqrt{2k^2-1},
\ee
and  the dn solutions are periodic with the periods $L_\textrm{dn}/A_\pm$, 
where
\be
\label{Ldn}
L_\textrm{dn}=L_\textrm{dn}(k)= 2K(k)\sqrt{2-k^2}.
\ee
In Eqs.\eqref{Lcn} and \eqref{Ldn},  $K$
is the complete elliptic integral of the first kind. As $k$ varies
from $1/\sqrt{2}$ to $1$, $L_{\textrm{cn}}$ grows, monotonically,
from $0$ to infinity. Less obvious fact is that $L_{\textrm{dn}}$ is
also a monotonically growing function -- growing from
$\sqrt{2}\pi$ to infinity as $k$ varies from $0$ to $1$.
(See the Appendix.) Therefore, for the given $h$ and $\gamma$ the period of the cnoidal
wave can be used as its third parameter, in lieu of the elliptic
modulus $k$.

As $k \to 1$, the periodic solutions
$\Psi_+^{\textrm{cn}}$ and $\Psi_+^{\textrm{dn}}$ approach the
$\psi_+$  soliton, while the $\Psi_-^{\textrm{cn}}$ and $\Psi_-^{\textrm{dn}}$ tend to the
soliton $\psi_-$  
\cite{B}.

\subsection{Linearised problem}
To examine the stability of cnoidal
waves~(\ref{sol1})--(\ref{sol2}), we write
$\psi(x,t)=\Psi_\pm(x)+\delta\psi(x,t)$, where $\Psi_\pm(x)$ is
the stationary solution in question, and linearize in
$\delta\psi$. Letting 
\[ 
\delta\psi(x,t)=\textrm{e}^{ -i \theta_\pm} \left[U(x,t)+i V(x,t)\right] 
\] 
yields
\be
\label{sesopr} 
U_T=\mathcal L_0V,\quad
-V_T-2 \tilde\gamma V=\mathcal L_1U,
\ee
where the operators $\mathcal L_{0,1}$ are given by
\begin{subequations} \label{lin_tri}
\be
\label{2.7}
 \mathcal L_0=-\frac{\textrm{d}^2}{\textrm{d}X^2}+1\mp\mathcal{E}-2q^2
\ee
and
\be
\label{2.8} 
\mathcal L_1=-\frac{\textrm{d}^2}{\textrm{d}X^2}+1-6q^2. 
\ee 
\end{subequations}
In  \eqref{sesopr},  $T=A_\pm^2t$ is the scaled
time variable, and in \eqref{2.7}-\eqref{2.8},  $X=A_\pm x$ is the
scaled spatial coordinate.
In \eqref{sesopr}-\eqref{lin_tri} we have scaled
the damping coefficient and introduced a  parameter $\cal E$
which measures the driving-damping difference:
\be
\label{varepsilon} \tilde\gamma=\gamma/A_\pm^2, \quad\mathcal
E=2\sqrt{h^2-\gamma^2}/A_\pm^2.
 \ee 
The notation $q$ stands for $q_\textrm{cn}$ or
$q_\textrm{dn}$ [Eq.~(\ref{qcn}) or Eq.~(\ref{qdn})] depending on
whether we linearize about a cn or dn solution. The top sign in
front of $\cal E$ in Eq.~(\ref{2.7}) corresponds to the
$\Psi_+^{\textrm{cn}}$ and $\Psi_+^{\textrm{dn}}$ solutions; the
bottom sign selects the   $\Psi_-^{\textrm{cn}}$ and
$\Psi_-^{\textrm{dn}}$ cnoidal waves. The cnoidal wave is deemed unstable
provided Eqs.~(\ref{sesopr}) have solutions growing faster than
$\textrm{exp}(\tilde\gamma T)$ in time.

We do not impose any periodicity requirements on $U$ and $V$.  All
we require is that $U(X,T)$ and $V(X,T)$ be bounded on the whole
line $-\infty<X<+\infty$.

We call $\Lambda$ a point of spectrum of an operator $\mathcal
L$ if the equation $\mathcal L y=\Lambda y$ has a solution
$y(X)$ bounded for all $X$, $\infty<X<\infty$.
The spectrum of  the operators (\ref{2.7}) and (\ref{2.8}) can be found exactly.
Consider first the cn solutions, that is, assume that $q(X)$ in
Eqs.~(\ref{2.7})--(\ref{2.8}) is given by Eq.~(\ref{qcn}). Then,
defining $\xi=X/\sqrt{2k^2-1}$, we obtain:
\be
\label{L0dn} 
\mathcal L_0=\frac{\frak{L}^{(1)}-1}  {2k^2-1}                       \mp\mathcal{E}, \ee
and 
\be\label{L1dn} \mathcal L_1=\frac {\frak{L}^{(2)}-1-2k^2}{2k^2-1}, \ee 
where
\[
\frak{L}^{(\ell)}= -\frac{\textrm{d}^2}{\textrm{d}\xi^2} + \ell (\ell+1) k^2
\textrm{sn}^2(\xi,k)
\]
is the  $\ell$-gap Lam\'e
operator~\cite{Arscott}.
Therefore the spectra of ${\mathcal L}_0$ and ${\mathcal L}_1$
result from the spectra of $\frak{L}^{(1)}$ and $\frak{L}^{(2)}$ (given in \cite{Arscott})
 by shift and scaling.

Thus, the
spectrum of $\mathcal L_0$ consists of a finite band
\be\label{211}
\Lambda \in 
\left[
\frac{k^2-1}{2k^2-1}\mp\mathcal{E}, \ \mp\mathcal{E}
\right], 
\ee
and  a semi-infinite band,
 \be\label{212}
 \Lambda \in  \left[
\frac{k^2}{2k^2-1}\mp\mathcal{E}, \  \infty
\right). \nonumber\ee
The spectrum of $\mathcal L_1$ comprises two finite bands \[
 \left[
-1-\frac{2\sqrt{1-k^2+k^4}}{2k^2-1}, \ - \frac{2k^2}{2k^2-1}
\right], \quad 
\left[  0, \ \frac{2(1-k^2)}{2k^2-1}\right],
\] 
and a semi-infinite band
\[
\left[
-1+\frac{\sqrt{1-k^2+k^4}}{2k^2-1}, \ \infty\right).\]

In the case of the dn solutions~(\ref{sol2}) the spectrum
structures are similar. 
Namely, assuming that $q(x)$ is given by
Eq.~(\ref{qdn}) and defining $\xi=X/\sqrt{2-k^2}$, 
the operators $\mathcal L_0$ and $\mathcal L_1$ can be expressed as 
\be
\label{L0d} 
\mathcal L_0=\frac{\frak{L}^{(1)}-k^2}  {2-k^2} \mp\mathcal{E} \ee
and 
\be\label{L1d} \mathcal L_1=\frac {\frak{L}^{(2)}-2-k^2}{2-k^2}, 
\ee 
respectively.
Accordingly, 
the spectrum of $\mathcal L_0$ comprises two
bands, 
\be
\left[
  \mp\mathcal{E}, \
\frac{1-k^2}{2-k^2}\mp\mathcal{E}
\right],
\quad
\left[ \frac{1}{2-k^2}\mp\mathcal{E}, \  \infty \right)
\label{215},
\ee
while the spectrum of $\mathcal L_1$ consists of three:
\begin{align}
\left[
 -1-\frac{2\sqrt{1-k^2+k^4}}{2-k^2}, \
 \frac{-2}{2-k^2}\right], \quad
\left[
-\frac{2(1-k^2)}{2-k^2}, \
0
\right], \nonumber \\
\left[
-1+\frac{\sqrt{1-k^2+k^4}}{2-k^2}, \ \infty \right). \label{216}
\end{align}
In Eqs.~(\ref{211})--(\ref{216}) the top sign in front of
$\mathcal{E}$ pertains to the $\Psi_+$ solutions and the bottom
sign  to $\Psi_-$.

The spectrum of the $\psi_+$ and $\psi_-$ solitons~\cite{B} is
recovered be sending $k\to 1$. In this case, each finite band
collapses into a discrete eigenvalue. As a result, the spectrum of
$\mathcal L_0$ consists of a single discrete eigenvalue
$\Lambda_0=\mp\mathcal{E}$ (and a continuum of values
$\Lambda\ge1\mp\mathcal{E}$) whereas the spectrum of $\mathcal
L_1$ includes two discrete eigenvalues, $\Lambda_0=-3$ and
$\Lambda_1=0$ ( and a continuum $\Lambda\ge 1$).

\subsection{Stability eigenvalues and symplectic  eigenvalue problem} 

We assume that the linear system  \eqref{sesopr} has 
separable solutions of the form
\[
U(X,T) = {\rm Re} \, \left[ e^{\eta T} {\tilde u}(X) \right];
\quad
V(X,T) = {\rm Re} \, \left[ e^{\eta T} {\tilde v}(X) \right],
\]
where ${\tilde u}$, ${\tilde v}$ and $\eta$ are complex.
Here $\eta$ and ${\tilde u}$, ${\tilde v}$ are eigenvalues and 
eigenfunctions in the eigenvalue problem
\be
{\mathcal L_0} {\tilde v}= \eta {\tilde u},
\quad
{\mathcal L_1} {\tilde u}= -(\eta+ 2 {\tilde \gamma}) {\tilde v}.
\label{S1}
\ee 
We will be referring to this problem as the
``linearised eigenvalue problem", whereas $\eta$ will be called ``stability eigenvalues" below.

Making a substitution $({\tilde u}, {\tilde v})  \to (u,v)$ where $\lambda {\tilde v}= \eta v$,  ${\tilde u}=u$,
and
\be
\lambda^2=\eta(\eta+ 2 {\tilde \gamma}),
\label{S2} 
\ee
we transform \eqref{S1} to
\be
{\mathcal H} {\vec y} = \lambda J {\vec y},
\label{S3}
\ee
where
\be
\mathcal H=
\left( 
\begin{array}{cc}
{\mathcal L_1}  & 0 \\
0 & {\mathcal L_0}
\end{array} 
\right),
\label{S30}
\ee
and 
\[
J=
\left( 
\begin{array}{cc}
0  & -1\\
1 & 0
\end{array} 
\right),
\quad
{\vec y}=
\left( \begin{array}{c}  u \\ v
\end{array} \right).
\]
One advantage of this formulation is that eigenvalues $\lambda$ depend
on $h$ and $\gamma$ only in combination $\mathcal E$
(that is, only as $h^2-\gamma^2$);
thus a two-parameter eigenvalue problem is
reduced to a one-parameter problem.
Another merit is that the operator $J^{-1} \mathcal H$ is symplectic, that is, 
generates a hamiltonian flow. The spectrum of symplectic operators 
consists of pairs of opposite pure-imaginary values, real pairs and
complex  quadruplets. If $\lambda$ is a real or pure imaginary point
of spectrum, then $-\lambda$ is another one;  if a complex $\lambda$ is in the spectrum, 
then so are $-\lambda, \lambda^*$, and $-\lambda^*$ \cite{Arnold}.
 We will be referring to
the eigenvalue problem \eqref{S3} as the ``symplectic eigenvalue problem",
and $\lambda$'s as the ``symplectic eigenvalues".
Having found 
a symplectic eigenvalue $\lambda$,
 we can readily recover the corresponding growth rate ${\rm Re}\, \eta$
from \eqref{S2}:
\be
{\rm Re} \, \eta= -{\tilde \gamma} + {\rm Re} \, \sqrt{{\tilde \gamma}^2+ \lambda^2}.
\label{S7}
\ee
(Here we have kept the largest of the two growth rates.)

Before proceeding to the analysis of the symplectic spectrum,
three remarks are in order. 
First of all, we note that if the symplectic eigenvalues $\pm \lambda$ are real, 
the corresponding stability eigenvalues are real as well:
$\eta= -{\tilde \gamma} \pm \sqrt{{\tilde \gamma}^2+ \lambda^2}$.
Thus the occurrence of a real symplectic eigenvalue immediately implies instability
of the underlying cnoidal wave. This instability has monotonic growth.

 Second, since the potentials in the 
operators $\mathcal L_0$ and $\mathcal L_1$ are 
even, it is sufficient to consider only even and odd  eigenfunctions 
${\vec y}(X)$.

Finally, we note that the potentials of the operators $\mathcal L_{0,1}$ are periodic with the period 
$L_{\mathrm{cn}}$ in the case of the $\mathrm{cn}^{\pm}$ solutions
and $L_{\mathrm{dn}}$ in the case of the $\mathrm{dn}^{\pm}$ cnoidal waves. 
Therefore, the subspace of  periodic functions with period $L_{\mathrm{cn}}$
respectively $L_{\mathrm{dn}}$ is invariant under the action of the operators $\mathcal L_{0,1}$
in the case of $\mathrm{cn}^{\pm}$ respectively $\mathrm{dn}^{\pm}$
solutions. This implies that the eigenvalue problem \eqref{S3} 
is well posed on the subspace of periodic functions  with the corresponding
period. 
The subspace of antiperiodic functions, that is, functions satisfying 
${\vec y}(X+L_{\mathrm{cn},\mathrm{dn}})=-{\vec y}(X)$, 
is also  invariant. Accordingly, 
the eigenvalue problem \eqref{S3} 
is well posed on the subspace of antiperiodic functions.

\section{Instability to periodic and antiperiodic perturbations}
\label{periodic_perturbations}

To establish instability of a solution it is sufficient to
demonstrate its instability to a particular class of perturbations. 
In this section we will show that the cnoidal waves 
$cn^-$ and $dn^-$ are unstable w.r.t. {\it periodic\/} perturbations,
for any combination of the parameters $h$, $\gamma$ and 
 $k$. We will also show that 
the wave $dn^+$ is  unstable to {\it antiperiodic\/} perturbations --- 
in some region of the parameter space. 

\subsection{Instability of $\mathrm{cn}^-$}
In the case of the cnoidal wave $cn^-$,
we will show that the symplectic eigenvalue problem \eqref{S3} has real eigenvalues $\lambda$
associated with the eigenfunctions satisfying  boundary conditions of the third kind,
\be
u_X\left( 0 \right) = u\left( \frac{L}{2} \right)=0,
\quad
v_X\left( 0 \right) = v\left( \frac{L}{2} \right)=0.
\label{Di}
\ee
Here $L=L_{\mathrm{cn}}$ is the half-period of the cnoidal wave, defined by Eq.\eqref{Lcn}. 
Due to the boundary condition at the origin, the eigenfunction ${\vec y}(X)$ has to be even.

Let $\mathcal L_0$ be defined by Eq.\eqref{2.7} with $q=q_{\rm cn}$ and 
consider an eigenvalue problem 
\be
\mathcal L_0 y=\Lambda y
\label{S4}
\ee
with the mixed boundary
conditions
\be
 y_X(0)=0; \quad y\left(\frac{L}{2}\right)=0.
 \label{F1}
  \ee 
One eigenvalue of $\mathcal L_0$ is $\Lambda_1=\mp \mathcal E$; it is
associated with the eigenfunction 
\be
\label{44}
y_1(X)=\textrm{cn}\left(X/\sqrt{2k^2-1}, \, k\right).
\ee 
The eigenfunction does not have zeros inside the interval $(0,L/2)$;
therefore $\Lambda_1$ is the lowest eigenvalue  of 
the regular Sturm-Liouville problem (\ref{S4})+(\ref{F1}).
If we are considering the
linearization about the solution $\textrm{cn}^-$, the
eigenvalue  $\Lambda_1=+\mathcal E$ is strictly positive.
Hence the operator $\mathcal L_0$, defined by the differential
expression (\ref{2.7}) and boundary conditions (\ref{F1}), is
positive definite. On the subspace of functions satisfying
(\ref{Di}), the system \eqref{S3} can be written in the form
\be
{\mathcal L_1} u= - \lambda^2 {\mathcal L_0}^{-1} u,
\label{F4}
\ee
where $\mathcal L_1$ is symmetric and $\mathcal L_0^{-1}$ symmetric and positive definite.
The smallest eigenvalue $(-\lambda^2)_0$ of the generalised eigenvalue problem \eqref{F4}
is given by the minimum of the Rayleigh quotient
\be
(-\lambda^2)_0 = \min_{u} \frac{\langle u|\mathcal L_1 |u\rangle } {\langle u|\mathcal L_0^{-1} |u\rangle},
\label{Ray}
\ee
where the minimum is evaluated over all functions $u(X)$ satisfying the
boundary conditions \eqref{Di}. 
Here the scalar product is defined by
\[
\langle u | v \rangle = \int_0^{L/2} u(X) v(X) dX.
\]

Turning to the operator $\mathcal L_1$, we note that it has a negative eigenvalue
 $\mu=-2k^2/(2k^2-1)$ with the eigenfunction
\be
\label{45}
z_1(X)=\textrm{cn}\left(\frac{X}{\sqrt{2k^2-1}},k\right)\textrm{dn}\left(\frac{X}{\sqrt{2k^2-1}},k\right)
\ee 
which satisfies the boundary condition (\ref{F1}). Therefore,
the quadratic form $\langle u|\mathcal L_1|u \rangle$ attains
negative values on the space of functions with the boundary
condition (\ref{F1}). Eq.\eqref{Ray} implies then that the smallest
eigenvalue $-\lambda^2$ of the  eigenvalue problem \eqref{F4}
is negative and hence the vector eigenvalue problem \eqref{S3} has a real eigenvalue $\lambda$.
By \eqref{S7} we conclude, 
eventually, that there are perturbations with positive growth rates ${\rm Re} \, \eta$.

Since the eigenfunction ${\vec y}(X)$ is even, ${\vec y}(L/2)=0$ implies
${\vec y}(-L/2)=0$. On the other hand, the derivative ${\vec y}_X(X)$
is odd;  thus we have
\be
{\vec y} \left( -\frac{L}{2} \right)= -{\vec y} \left( \frac{L}{2} \right), \quad
{\vec y}_X \left( -\frac{L}{2} \right)= -{\vec y}_X \left( \frac{L}{2} \right).
\label{anti}
\ee
Considering Eq.\eqref{S3} as a system of four first-order equations 
with $L$-periodic coefficients, Eq.\eqref{anti} implies  that its real
 solution $(u, u_X, v, v_X)$ is antiperiodic on the interval $(-L/2,L/2)$ --- that is, has the Floquet multiplier $-1$.
By the Floquet theorem, 
this solution (and hence the eigenfunction $\vec y$) is periodic with the period
$2L$.

We
conclude that the cnoidal wave $\textrm{cn}^-$ is unstable against periodic perturbations
of period $2L_{\rm cn}$ (the period of the wave), for any combination of 
$h$, $\gamma$ and $k$.

\subsection{Instability of $\mathrm{dn}^-$}

The case of the wave ${\mathrm dn}^-$ is different from the 
situation in the previous section in that the operators $\mathcal L_0$ and $\mathcal L_1$
have potentials periodic with the period of the cnoidal  wave (rather than
with the half period).
Nevertheless, the above proof remains in place, with some 
modifications.

In this case, the spectrum of the operator $\mathcal L_0$ lies on the positive
semiaxis. Therefore, the operator $\mathcal L_0$ is positive definite for any 
boundary conditions imposed on $y(X)$ in \eqref{S4}, in particular for the 
boundary conditions 
\be
 y_X(0)=0; \quad
  y_X \left(\frac{L}{2}\right)=0.
 \label{F2}
  \ee
   Therefore, on the subspace of functions satisfying the Neumann boundary conditions
 \be
u_X\left( 0 \right) = u_X\left( \frac{L}{2} \right)=0,
\quad
v_X\left( 0 \right) = v_X\left( \frac{L}{2} \right)=0
\label{Dii}
\ee  
 the system \eqref{S3} can be written in the form  \eqref{F4}
 where $\mathcal L_1$ is symmetric and $\mathcal L_0^{-1}$ symmetric and positive definite.
 As before, the minimum eigenvalue of \eqref{F4} is given by the 
 minimum of the Rayleigh quotient \eqref{Ray}
 where this time the minimum is evaluated over all $u(X)$ satisfying the 
 Neumann conditions \eqref{Dii}.

The operator $\mathcal L_1$ has 
a negative eigenvalue 
\[
\mu_1= -1- \frac{ \sqrt{k^4+1-k^2}}{2-k^2}
\]
with an eigenfunction 
\[
z_1(X)= dn^2 \left( \frac{X}{\sqrt{2-k^2}} \right)+ 
\frac{k^2+\sqrt{k^4-k^2+1} }{3}
\]
which satisfies \eqref{F2}.
Therefore, the minimum of the Rayleigh quotient is negative
on the space of $u$ satisfying \eqref{F2}; 
the vector eigenvalue problem \eqref{S3} has a real eigenvalue $\lambda$,
and the growth rate \eqref{S7} is positive.

Recalling that the eigenfunction ${\vec y}(X)$ can be chosen either even or odd, the
condition
 ${\vec y}_X(0)=0$ selects the even function.
The boundary conditions \eqref{Dii} imply then that 
${\vec y}_X(-L/2)= 0$
and so 
the eigenfunction ${\vec y}(X)$ is $L$-periodic:
\be
{\vec y}\left( -\frac{L}{2} \right)= {\vec y} \left( \frac{L}{2} \right), \quad
{\vec y}_X \left( -\frac{L}{2} \right)= {\vec y}_X \left( \frac{L}{2} \right).
\label{proti}
\ee

We have thus established instability of the $dn^-$ solution
against perturbations periodic with the period $L_{\rm dn}$,
the period of the $dn$ cnoidal wave. The $dn^-$ wave is unstable 
for any choice of the parameters $h$ and $\gamma$, and any $k$.

\subsection{Instability of $\textrm{dn}^{+}$:  antiperiodic perturbations}

So far we have demonstrated the instability of the $\mathrm{cn}^-$ and $\mathrm{dn}^-$
waves, with the unstable perturbations  exhibiting a monotonic growth.
Another solution that turns out to be prone to the instability of a similar type is
the $\textrm{dn}^{-}$ wave; however this time our proof will only be valid
in a part of the $(h, \gamma, k)$ parameter space.

In the case of the $\textrm{dn}^{+}$ wave, the  operator $\mathcal L_0$ has an eigenvalue
 $\tilde\Lambda_1=1/(2-k^2)-\mathcal E$ with an eigenfunction
\[
\tilde y_1(X)=\textrm{sn}\left(\frac{X}{\sqrt{2-k^2}},k\right).
\]
The eigenfunction $\tilde y_1$ 
satisfies mixed boundary conditions 
\be
y(0)=0, \quad y_X \left( \frac{L}{2} \right)=0
\label{F3}
\ee
and does not have zeros inside the interval $(0, L/2)$. Hence the eigenvalue $\tilde\Lambda_1$
is the smallest eigenvalue of $\mathcal L_0$ under the boundary conditions \eqref{F3}. 

The eigenvalue $\tilde\Lambda_1$ is positive and the operator $\mathcal L_0$
is positive definite in two adjacent parameter regions. One parameter region is
 \be 
 \label{52}
  h<\sqrt{\frac{1}{9}+\gamma^2},
\ee 
with $k$ taking any values between $0$ and $1$. The second region is
\begin{subequations}
\label{53}
\be
\sqrt{\frac{1}{9}+\gamma^2}< h <  \sqrt{1+\gamma^2},
\ee
with the elliptic modulus being bounded from below:
\be
k^2 >\frac{3}{2}-\frac{1}{2}\frac{1}{\sqrt{h^2-\gamma^2}}.
\ee
\end{subequations} 
We now assume that the parameter vector $(h, \gamma, k)$ 
lies in one of the above two regions. 

The operator
$\mathcal L_1$ has a negative eigenvalue 
$2(k^2-1)/(2-k^2)$ with an eigenfunction satisfying the boundary conditions 
\eqref{F3}:
\[
\tilde
z_1(X)=\textrm{sn}\left(\frac{X}{\sqrt{2-k^2}},k\right)\textrm{dn}\left(\frac{X}{\sqrt{2-k^2}},k\right).
\]
Therefore the minimum of the Rayleigh quotient \eqref{Ray} on the 
space of functions satisfying \eqref{F3} is negative and
 the symplectic eigenvalue problem 
 \eqref{S3} has a real eigenvalue $\lambda$.
 This means that the $\mathrm{dn}^+$ wave is unstable to perturbations satisfying 
  \[
u \left( 0 \right) = u_X\left( \frac{L}{2} \right)=0,
\quad
v \left( 0 \right) = v_X\left( \frac{L}{2} \right)=0.
\]

The boundary condition ${\vec y}(0)=0$
singles out the odd eigenfunction; hence it satisfies
the antiperiodicity conditions on the interval $(-L/2,L/2)$:
\[
{\vec y}\left( -\frac{L}{2} \right)= -{\vec y} \left( \frac{L}{2} \right), \quad
{\vec y}_X \left( -\frac{L}{2} \right)= -{\vec y}_X \left( \frac{L}{2} \right).
\]

We conclude that when  $h,\gamma$ and $k$ belong to the region \eqref{52}+\eqref{53},
the cnoidal wave $\mathrm{dn}^+$ is unstable under perturbations of
 period twice the period of the wave.

\vspace*{3mm}
\section{The unperturbed nonlinear Schr\"odinger:  Stability of the $cn$ wave}
\label{CN_E0_pert}

When $h=\gamma$,  Eq.\eqref{varepsilon} gives  $\mathcal E=0$ and the spectrum 
of symplectic eigenvalues \eqref{S3} coincides with the spectrum 
of stability eigenvalues of the 
cnoidal wave of the undamped undriven NLS.

\subsection{Small eigenvalues: general setting}

In this subsection we will obtain small symplectic eigenvalues.

Factorising the eigenfunction ${\vec y}$ into a  periodic function $\vec {\mathcal Y}$ and an exponential, 
${\vec y}= {\vec {\mathcal Y}} e^{i \kappa X}$, 
the symplectic eigenvalue problem \eqref{S3} becomes
\be
\mathcal{H} {\vec {\mathcal Y}}= 2i \kappa {\vec {\mathcal Y}}_X - \kappa^2 {\vec {\mathcal Y}} + \lambda J {\vec {\mathcal Y}}.
\label{T1}
\ee
Without loss of generality,
the period of the function  $\vec {\mathcal Y}$ could be taken  equal to the period of the potential of the 
operator  $\mathcal H$:   $L_{\mathrm cn}$ in the case of the $cn$-wave, and  $L_{\mathrm dn}$ in the case of the $dn$-wave.
This will indeed be our choice in the $dn$ situation. However, in the case
of the $cn$ wave, it is convenient to regard $\vec {\mathcal Y}$ as a $2L_{\mathrm cn}$-periodic function --- 
that is, choose the period of $\vec {\mathcal Y}$ to coincide with the period of the cnoidal wave. 
This convention is equivalent to the previous one, and is equally general.

The symplectic spectrum includes a four-fold zero eigenvalue;
associated with these are two periodic eigenvectors and two generalised eigenvectors.
When $\kappa$ is small,  we expand the 
eigenvalues and eigenfunctions in powers of $\kappa$:
\begin{align}
{\vec {\mathcal Y}}= {\vec {\mathcal Y}}_0+ \kappa {\vec {\mathcal Y}}_1
+ \kappa^2 {\vec {\mathcal Y}}_2+..., \nonumber \\
\lambda= \lambda_1 \kappa + \lambda_2 \kappa^2 +...
\label{T1A}
\end{align}
Substituting in \eqref{T1} we equate coefficients of like powers of $\kappa$. 

The coefficient of $\kappa^0$ gives ${\mathcal H} {\vec {\mathcal Y}}_0=0$.
The general solution is of  the form 
\be
{\vec {\mathcal Y}}_0= 
\left( \begin{array}{c} \mathcal{A} q_{X} \\ \mathcal{B} q \end{array} \right),
\label{T2}
\ee
where $\mathcal A$ and $\mathcal B$ are two arbitrary constants. Next, the order $\kappa^1$ produces
\be
{\mathcal H} {\vec {\mathcal Y}}_1= 2i \partial_X {\vec {\mathcal Y}}_0+ \lambda_1 J {\vec {\mathcal Y}}_0,
\ee
or, componentwise, 
\bea 
{\mathcal L}_1 u_1= 2i \mathcal{A} \partial_X^2 q- \lambda_1 \mathcal{B} q,  \label{T3} \\
{\mathcal L}_0 v_1= (2i\mathcal{B}+ \lambda_1 \mathcal{A}) q_X, \label{T4}
\eea
where $u_1$ and $v_1$ are the top and bottom components of the 
vector ${\vec {\mathcal Y}}_1$: ${\vec {\mathcal Y}}_1=(u_1, v_1)^T$. We note that 
\begin{align*}
{\mathcal L}_1 \partial_X (Xq)=-2q, \\
{\mathcal L}_1 (Xq_X)= -2 \partial_X^2 q, \\
{\mathcal L}_0 (Xq)= -2q_X;
\end{align*}
therefore one solution to Eq.\eqref{T3} is 
\be
{\widetilde u_1}= -i \mathcal{A} Xq_X+ \frac{\lambda_1}{2} \mathcal{B} (Xq)_X, \label{T5}
\ee
and one solution to \eqref{T4} is 
\be
{\widetilde v_1}= -\frac12 (\lambda_1\mathcal{A} + 2i\mathcal{B}) Xq. \label{T6}
\ee
However neither of these solutions is periodic.

\subsection{$cn$ wave}

From this point on, our analysis depends on which periodic solution we consider.
We start with the $cn$ wave. In this case both components of the zero-order 
approximation 
${\vec {\mathcal Y}}_0$ are $2L$-periodic, and we will attempt to construct
its perturbation ${\vec {\mathcal Y}}$ with the same period. Here
 $L$ is our short-hand notation 
for $L_{\mathrm cn}$. 

In order to obtain a periodic $u_1$, we can add to ${\widetilde u_1}$
a multiple of $\partial_kq$, the 
nonperiodic homogeneous solution of equation \eqref{T3}:
\begin{align}
  \partial_k q(x)=
  \textrm{sn} (\xi,k) \textrm{dn} (\xi,k) \nonumber \\
  \times 
 \left[ \frac
 {E(\textrm{am}  \,\xi,k)}{(1-k^2)(2k^2-1)}
 + \frac{X}{(2k^2-1)^2}
 \right]  \nonumber   \\
 -\frac{ \textrm{cn} (\xi,k)} {(2k^2-1)^{1/2}}
 \left[\frac{1}{2k^2-1}+
 \frac{k^2}{1-k^2}   \textrm{sn}^2 (\xi,k)
 \right].
 \label{qk}
 \end{align}
 In \eqref{qk},  $\xi= X/\sqrt{2k^2-1}$  and $E(\varphi,k)$ is the incomplete 
elliptic integral of the second kind:
\[
E(\varphi,k)= \int_0^\varphi \sqrt{1-k^2 \sin^2 \theta} d \theta.
\]
 The resulting nonhomogeneous solution 
\be
u_1= -i \mathcal{A} Xq_X+ \frac{\lambda_1}{2} \mathcal{B} (Xq)_X +C\partial_kq
\label{T7}
\ee
is even and so the periodicity condition $u_1(L)=u_1(-L)$ is satisfied
automatically,  for any $C$;  therefore,
one only needs to satisfy $\partial_X u_1(L)=\partial_X u_1(-L)$. 
For the odd $\partial_X u_1(X)$, this reduces to
\be
\partial_X u_1(L)=0. 
\label{T8}
\ee
Substituting \eqref{T7} in \eqref{T8} we find
\[
C= \left( \lambda_1 \mathcal{B}-i \mathcal{A} \right) \frac{L}{L_k},
\]
where $L_k= \partial_k L$. 

In a similar way, we take
\be
v_1= -\left( \frac12 \lambda_1 \mathcal{A}+i \mathcal{B} \right) Xq+ D z,
\label{T9}
\ee
where $z$ is  the nonperiodic homogeneous solution of \eqref{T4},
\begin{align*}
z(X)= \xi \mathrm{cn} 
\left( \xi, k \right) \\
+ \frac{1}{1-k^2} \left[
\mathrm{sn} \left( \xi, k \right) \mathrm{dn} \left( \xi, k \right)
- 2\mathrm{cn}  \left( \xi, k \right)
E(\mathrm{am}
 \xi, k) \right].
\end{align*}
Eq. \eqref{T9} is odd and its derivative $\partial_X v_1$ is even;
hence the periodicity condition 
$\partial_X v_1(L)= \partial_X v_1(-L)$ is in place
and we only need to make sure that $v_1(L)=v_1(-L)$. This amounts to
\be
v_1(L)=0.\label{T20}
\ee
Substituting Eq.\eqref{T9} in \eqref{T20},
the constant $D$ is identified:
\[
D=- \left(  \lambda_1 \mathcal{A}+i \mathcal{B} \right) 
\frac{K}{K_k}.
\]

\begin{widetext}

\begin{figure}
\includegraphics[width = \linewidth]{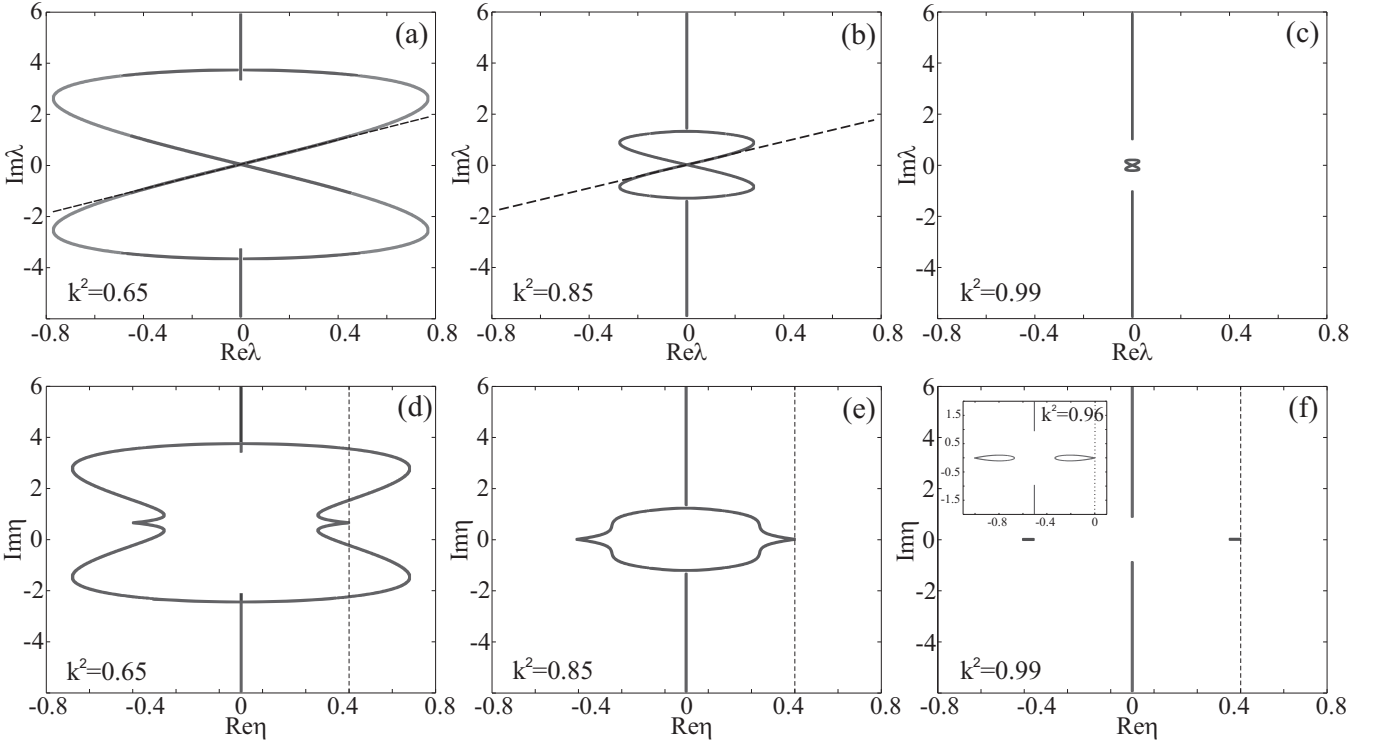}
\caption{\label{fig_cn}
Spectra of the $\textrm{cn}$ wave with $h=\gamma$. 
Top row: the spectrum of the symplectic operator $J^{-1} \mathcal{H}$ with $\mathcal E=0$.
This is essentially the spectrum of the cn-wave solution of the unperturbed NLS.
The dashed line in (a) and (b) makes the angle $\phi$ to the horizontal axis, where
$\phi$ is as in \eqref{T11}.
Bottom row: the corresponding spectrum of the linearised operator \eqref{S1}
with $h=\gamma$. In these plots, $\gamma=0.5$.
}
\end{figure}

\end{widetext}

At the order $\kappa^2$ we obtain
\begin{align}
\mathcal{L}_1u_2= 2i \partial_X u_1-u_0-\lambda_2 v_0 - \lambda_1 v_1,  \label{T100A} \\
\mathcal{L}_0 v_2= 2i \partial_X v_1-v_0+ \lambda_2 u_0+ \lambda_1 u_1, \label{T100}
\end{align}
where $(u_2,v_2)^T= {\vec {\mathcal Y}}_2$.
The solvability conditions for these two equations are, respectively: 
\begin{align}
2i \langle q_X | \partial_X u_1\rangle - \mathcal{A} \langle q_X | q_X\rangle- \lambda_1 \langle q_X | v_1\rangle=0, \label{T101}\\
2i\langle q | \partial_X v_1\rangle - \mathcal{B} \langle q | q\rangle + \lambda_1 \langle q| u_1\rangle=0.
 \label{T102}
\end{align}
Here
\[
\langle f | g\rangle= \int_{-L}^L f(X) g(X) dX.
\]
Substituting for $u_1$ and $v_1$ we obtain a system of linear 
equations
\be
\mathcal{M} 
\left(
\begin{array}{c} \mathcal{A} \\ \mathcal{B}
\end{array}
\right) =
\left(
\begin{array}{cc}
m_{11} \lambda_1^2 +n_{11} & im_{12} \lambda_1 \\
im_{21} \lambda_1& m_{22} \lambda_1^2 +n_{22}
\end{array} \right)
\left(
\begin{array}{c} \mathcal{A} \\ \mathcal{B}
\end{array}
\right)=0,
\label{M_cal}
\ee
where the elements of the matrix $\mathcal{M}$ are defined by
\begin{align} 
m_{11}= \frac12 \langle q_X | w \rangle, 
\quad
n_{11}= \langle q_X | 2p_X-q_X \rangle, \nonumber \\
m_{12}= \langle q_X | q_X+p_X+w \rangle, \quad
m_{21} = -\langle q  | p +w_X \rangle,   \nonumber \\
m_{22}= \frac12 \langle q| q+p\rangle, 
\quad
n_{22}= \langle q| 2w_X-q\rangle.
\label{Mel}
\end{align}
Here
\[
p= Xq_X+\frac{L}{L_k} q_k,
\quad
w= Xq+ \frac{K}{K_k} z.
\]
The integrals are easily evaluated:
\begin{align*}
m_{11}=
-\frac{k(1-k^2)}{2(2k^2-1)}
\frac{(LN)_k}{N}, \quad
n_{11}= -\frac{2k}{(2k^2-1)^2} \frac{L^2}{L_k}, \\
m_{12}=m_{21}= \frac{8}{(2k^2-1)(1-k^2)}
\frac{E(E-K)}{K_k L_k},  \\
m_{22}=\frac12 \frac{(LN)_k}{L_k}, \quad
n_{22} = \frac{4k}{(2k^2-1)} \frac{K^2}{K_k}, 
\end{align*}
where
\[
N= \int_{-L}^L q^2(X) dX= \frac{4(1-k^2)}{\sqrt{2k^2-1}} K_k
\]
and $L$ given by Eq.\eqref{Lcn}.
Setting $\mathrm{det} \, \mathcal{M}=0$ yields
a  biquadratic equation
\be
a_0 \lambda_1^4 + b_0 \lambda_1^2 +c_0=0, \label{biqua}
\ee
where
\[
a_0= \left[ (LN)_k \right]^2, \quad 
b_0=\frac{8kL^2N^2}{2k^2-1} \frac{L}{N},   \]
and \[
c_0=\frac{16 k^2 L^4}{(2k^2-1)^2}.
\]
The roots of the equation are given by $\pm \lambda_1, \pm \lambda_1^*$, where
\be
\lambda_1= \rho e^{i \phi}, \quad
\cot \phi= \frac{\sqrt{2k^2-1}}{k} \frac{K-E}{E}.
\label{T11}
\ee
The roots have nonzero real parts; this implies that the $\mathrm{cn}$-solution of the 
unperturned nonlinear Schr\"odinger equation is unstable for 
all $k$. 
This fact is known in literature.

The eigenvalues $\lambda=\lambda_1 \kappa+O(\kappa^2)$,  with small $\kappa$,
lie on four rays emanating out of the origin on the complex-$\lambda$ plane. 
The full spectrum of symplectic
eigenvalues, obtained numerically, is displayed in the top row of Fig.\ref{fig_cn}.
The result \eqref{T11} for the dominant behaviour of the eigenvalues near the 
origin is shown by the dashed line. The analytical result is seen to accurately reproduce
the numerically computed eigenvalues.
The four rays bend and join, pairwise, forming 
an eight-shaped curve 
centred at the origin (Fig.\ref{fig_cn} (a-c)).
In addition, the spectrum fills
the 
imaginary axis of $\lambda$ (with a gap).

\subsection{The unperturbed nonlinear Schr\"odinger:  Stability of the $dn$ wave}

In the case of the dn wave,  the eigenvalue problem \eqref{T1}
and expansion \eqref{T1A} remain in place. The null eigenfunction ${\vec {\mathcal Y}}_0$
is periodic with the period of the $dn$ wave: ${\vec {\mathcal Y}}_0(X+L)= {\vec {\mathcal Y}}_0(X)$,
where $L=L_{\mathrm dn}$ is as in \eqref{Ldn},
and we will assume that the eigenfunction ${\vec {\mathcal Y}}$
satisfies the same boundary conditions.

The $\epsilon^1$-corrections $u_1$ and $v_1$ are still given by
equations \eqref{T7} and \eqref{T9}, respectively, where the homogeneous
solutions are, this time,
\begin{align*}
\partial_k q= \frac{k  \, \mathrm{sn} (\xi, k) \mathrm{cn} (\xi,k) }{(2-k^2)^{1/2}} 
\left[\frac{E(\mathrm{am} \xi)}{1-k^2}- \frac{X}{(2-k^2)} \right] \nonumber \\
+ \frac{k \, \mathrm{dn} (\xi,k)}{(1-k^2)(2-k^2)} 
\left[ (2-k^2) \mathrm{cn}^2 (\xi,k)-1 \right]
\end{align*}
and 
\[
z(\xi)=k^2 \textrm{cn}(\xi, k)
\textrm{sn}(\xi,k)- 2\textrm{dn}(\xi,k)E(\textrm{am}\xi, k),
\]
where
\[
\xi=\frac{X}{\sqrt{2-k^2}}.
\]

Since the function 
\be
u_1(X)=-iA X q_X+\frac{\lambda_1}{2} B (Xq)_X + C q_k
\label{D1}
 \ee
is even, the only periodicity condition that needs to be verified, is 
\be
\partial_X u_1 \left( \frac{L}{2} \right)=0.
\label{D2}
\ee
Substituting \eqref{D1} in \eqref{D2}, the constant $C$ is evaluated to be 
\[
C= \left(  \lambda_1 \mathcal{B}-i \mathcal{A} \right) \frac{L}{L_k},
\]
where $L_k= \partial_k L$. On the other hand, the function 
\[
v_1= -\left( \frac12  A \lambda_1 + iB\right) Xq +Dz
\]
is odd; hence the periodicity condition reduces to
\[
v_1 \left( \frac{L}{2} \right)=0.
\]
This gives
\[
D= -\left(  \lambda_1 A+ iB \right) \frac{K}{E}.
\]

At the order $\kappa^2$ we obtain equations \eqref{T100A} -\eqref{T100},
with the solvability conditions \eqref{T101}-\eqref{T102}, where, this time, 
the scalar product is defined as
\[
\langle f | g\rangle= \int_{-L/2}^{L/2} f(X) g(X) dX.
\]
The elements of the matrix \eqref{M_cal} are given by Eqs.\eqref{Mel}
where, this time, the functions $p(X)$ and $w(X)$ are given by
\[
p=2X q_X+\frac{L}{L_k}q_k(X), \quad w=X q+\frac{K}{E} z(X).
\]
Substituting for $q$ we get
\begin{align*}
m_{11}=-\frac{k(1-k^2)}{4(2-k^2)^2}\frac{(LN)_k}{N}, 
\quad
 n_{11}=\frac{k^2}{(2-k^2)^2} \frac{L^2}{L_k}, \\
 m_{12}=m_{21}=
 - \frac{{k^\prime}^2}{(2-k^2)^2}
 \frac{L}{NL_k} \\
 \times
 \left(\sqrt{1-k^2} N\right)_k \left( \frac{L}{\sqrt{1-k^2}} \right)_k,
 \\
m_{22}=\frac{1}{2}\frac{(LN)_k}{L_k}, 
\quad
n_{22} =-\frac{1-k^2}{(2-k^2)^3}\frac{L^2}{N} , 
\end{align*}
where $L$ is given by \eqref{Ldn} and
\be
N=\int_{L/2}^{L/2}q^2(X)\textrm{d}X=\frac{2}{\sqrt{2-k^2}}E.
\label{Ndn}
\ee

Hence we find the coefficients of the biquadratic equation \eqref{biqua}:
\[
a_0=\left[ (LN)_k \right]^2, \quad c_0=
\frac{8 k^2}{(2-k^2)^3} L^4,
\]
and
\[
b_0=
-\frac{4}{k(2-k^2)}L N^2 \left( \frac{L}{N} \right)_k.
\]
The discriminant of the equation is
\[
\mathcal{D} =
\frac{4L^4}{k (2-k^2)^2}
({k^\prime}^2L_kN- N_k L) (L_k N- {k^\prime}^2 N_k L).
\]
Since $L_k>0$ and $N_k<0$, this is positive and so Eq.\eqref{biqua} has
two positive roots, $(\lambda_1^2)_a>0$ and $(\lambda_1^2)_b>0$.

\begin{widetext}

\begin{figure}
\includegraphics[width = \linewidth]{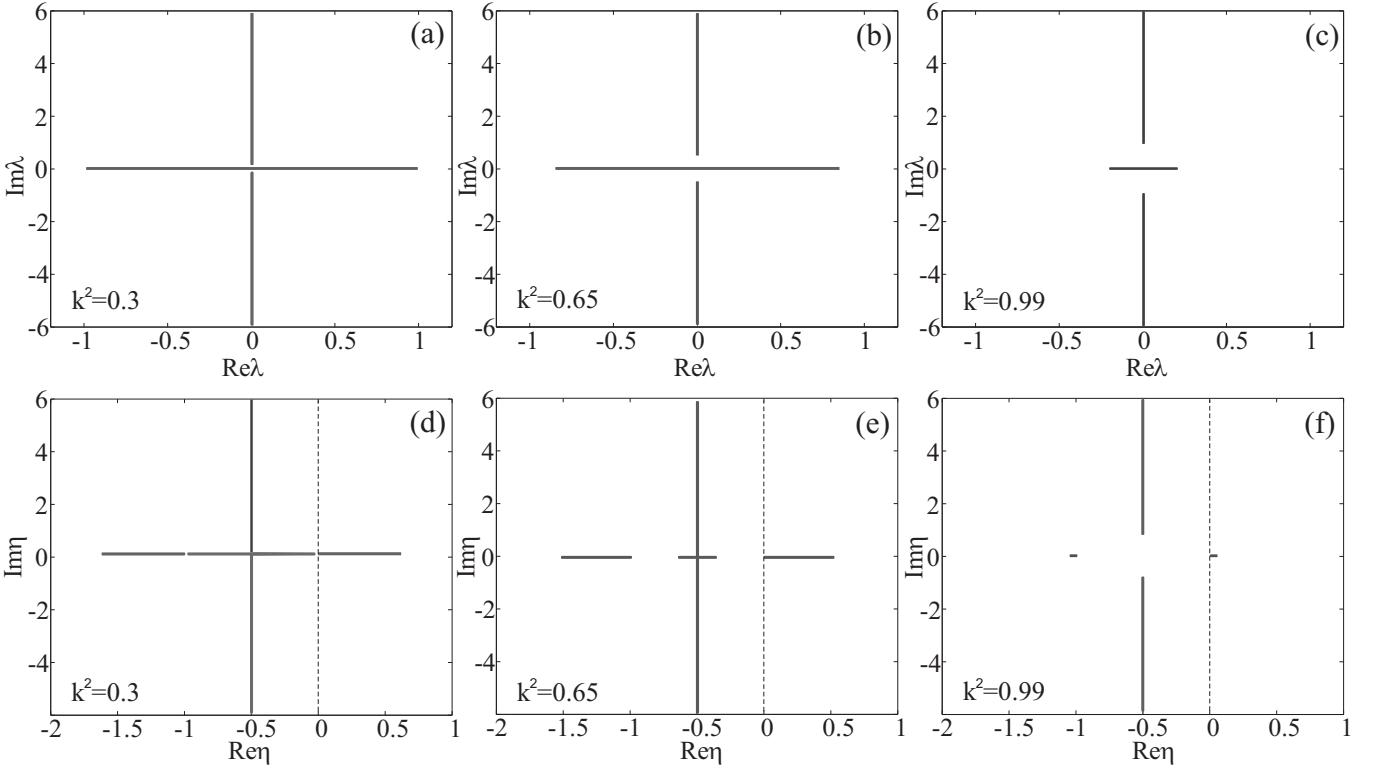}
\caption{\label{fig_dn}
Spectra of the $\textrm{dn}$ wave with $h=\gamma$. 
Top row: the spectrum of the reduced operator \eqref{S3} with $\mathcal E=0$.
Bottom row: the corresponding spectrum of the linearised operator \eqref{S1}
with $h=\gamma$. In these plots, $\gamma=0.5$. }
\end{figure}
\end{widetext}

The conclusion is that the $dn$-wave solution of the unperturbed NLS is unstable, for any $k$.
This fact is known to workers in the field. Furthermore, the real positive symplectic
eigenvalues $\lambda$ translate into
real positive growth rates $\eta$. This implies that the dn wave is also unstable as a 
solution of the damped-driven NLS with $h=\gamma$.

\section{The damped-driven $\mathrm{cn}$ waves with $h=\gamma$}

That  the $\textrm{cn}$ wave of the {\it unperturbed\/} NLS
is unstable for any $k$  does not mean, however,  that the {\it damped-driven\/} wave 
with $h=\gamma$ is necessarily unstable. In the damped-driven
situation, the stability is determined by the linearised eigenvalues $\eta$, not 
symplectic eigenvalues $\lambda$. Transforming $\lambda$ to $\eta$ 
by the rule \eqref{S2}, the symplectic spectrum shown in the top row of Fig.\ref{fig_cn} 
is mapped to the spectrum shown in the bottom row of the same figure.

\subsection{Stability to long-wavelength perturbations}

The symplectic spectrum includes a double zero eigenvalue
resulting from the phase invariance of the unperturbed NLS equation
(invariance w.r.t. constant and constant-velocity  phase rotations)
and another double zero resulting from its translation and Galilian invariances.
The map \eqref{S2} leaves only two eigenvalues at the origin: one resulting from 
the translation symmetry and the other one corresponding to the symmetry w.r.t. 
the velocity boosts. 

Next, for small $\kappa \ll {\tilde \gamma}$, the small-$\eta$ branch of the
map is simply $\eta= (2 {\tilde \gamma})^{-1} \lambda^2$.
Hence any ray $\lambda = e^{i \phi} r$ emanating 
out of  the origin on the $\lambda$-plane 
and making the angle $\phi$ to the real axis,
 is mapped to a ray making double that angle to 
 the real axis on the $\eta$-plane:
$\eta=e^{2i \phi} r'$. 
The branches of the spectral curve with $\mathrm{Re} \, \lambda>0$
emanate out of the origin as two rays making the angles
$\pm \phi$  to horizontal.
Eq.\eqref{T11}  implies that 
the value of $\cot \phi$ is smaller than 1   for all $k > \frac{1}{\sqrt{2}}$
(see the Appendix) and so $\phi$ lies between $\frac{\pi}{4}$ and $\frac{\pi}{2}$. 
These two rays (and their negatives) are mapped to rays with $\mathrm{Re} \, \eta <0$. Therefore, no instability 
of the damped-driven cnoidal waves with $h=\gamma$ 
is associated with
the neighbourhood of $\eta=0$, no matter what is the value of $k$.

\begin{figure}
\includegraphics[width = \linewidth]{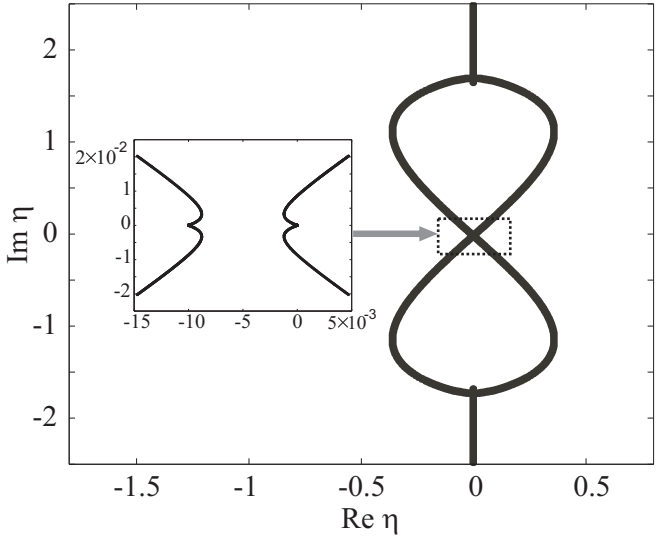}
\caption{\label{eight_inset}
Spectrum of linearised eigenvalues of the cn solution
with very small $\gamma$. (In this plot, $\gamma=0.005$ and $k^2=0.8$.)
On a large scale, the  spectrum of $\eta$ looks like the spectrum of $\lambda$;
however zooming in on the neighbourhood of the origin (dotted box,
enlarged in the inset) the shape typical for $\eta$-curves emerges.
}
\end{figure}

This conclusion is valid for {\it all\/} $\gamma$, including very small ones. A natural question,
therefore, is how the stable $\eta$-spectrum becomes unstable when $\gamma$ reaches zero;
that is, how can a diagram from the bottom row in Fig.\ref{fig_cn}
(two rays at the angle $2 \phi$ to horizontal)
evolve into the corresponding diagram in the top row
(two rays at the angle $\phi$ to horizontal).
The answer is that for small ${\tilde \gamma}$,
${\tilde \gamma} \ll \kappa$,  the 
 map \eqref{S2} reduces to $\eta=\lambda$.
 On a relatively large scale, namely on the scale $|\eta| \gg {\tilde \gamma}$, the spectrum
 of $\eta$ looks indistinguishable from the corresponding spectrum of 
 $\lambda$; it includes an eight-shaped curve centred at the origin.
 However if we zoom in on {\it very\/} small $\eta$ ($|\eta| \sim {\tilde \gamma}$),
 we will observe that the four rays do not reach the origin on the $\eta$-plane,
 but form a shape shown in the inset to Fig.\ref{eight_inset}. 
 As $\gamma \to 0$, the box shown in the inset shrinks to the origin.

\subsection{The  $\mathrm{cn}$ waves with $h=\gamma$: arbitrary perturbations}

\begin{figure}
\includegraphics[width = \linewidth]{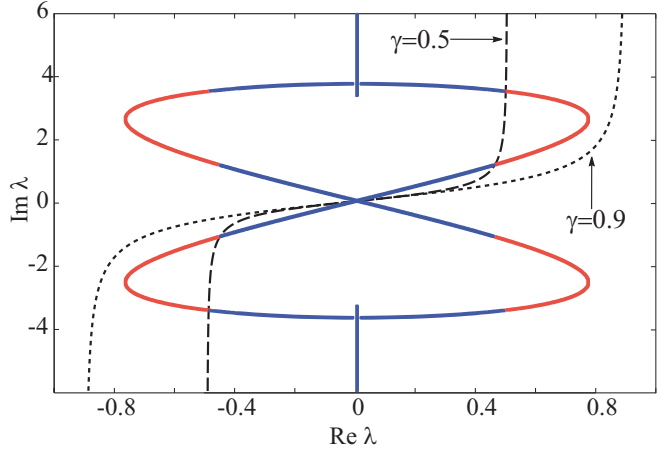}
\caption{\label{eight_asymp}
The symplectic spectrum of the cn wave (solid) and the curves \eqref{crs}
(dashed).
For small ${\tilde \gamma}$, the curves intersect the spectrum. The ``internal"
section of the spectrum corresponds to linearised eigenvalues with $\mathrm{Re} \, \eta<0$
whereas the part which lies outside the dashed curves is ``unstable": $\mathrm{Re} \, \eta>0$.
For large ${\tilde \gamma}$, the entire symplectic spectrum is stable.
In this plot, $k^2=0.65$.
}
\end{figure}

The periodic wave is unstable if the linearised-spectrum curve
crosses into the $\mathrm{Re\/} \, \eta>0$
half-plane. For the crossing points $\mathrm{Re\/} \, \eta=0$, Eq.\eqref{S2} gives
\be 
\mathrm{Im} \, \lambda = \pm \frac{\mathrm{Re} \, \lambda}{\sqrt{1- (\mathrm{Re} \, \lambda / {\tilde \gamma})^2}}.
\label{crs}
\ee
This is an equation for two curves on the $\lambda$-plane. 
These two curves  are straight lines at the origin, making $45^\circ$ to the horizontal.
As $\mathrm{Im} \, \lambda \to \pm \infty$, the curves are asymptotic to the 
vertical straight lines $\mathrm{Re} \, \lambda = \pm {\tilde \gamma}$
(see Fig.\ref{eight_asymp}). 

\newpage

For the given $h=\gamma$, the cnoidal wave is unstable if  the corresponding pair of 
curves \eqref{crs} intersects
the locus of the symplectic spectrum on the $\lambda$-plane.
Since the rays of the eight-shaped spectral curve make the angle  
 {\it greater\/} than $45^\circ$ to horizontal at the origin, 
the curves \eqref{crs} do not have to intersect the spectral curve.
They
will cross through the ``eight" only when ${\tilde \gamma}$  is large enough
(Fig.\ref{eight_asymp}). 
For some critical value ${\tilde \gamma}={\tilde \gamma_c}(k)$, the curves will just touch the ``eight".
This critical value defines the range of stability of the cn wave for the given $k$: the wave is stable 
when ${\tilde \gamma}> {\tilde \gamma_c}(k)$ and unstable otherwise.

\begin{figure}
\includegraphics[width = \linewidth]{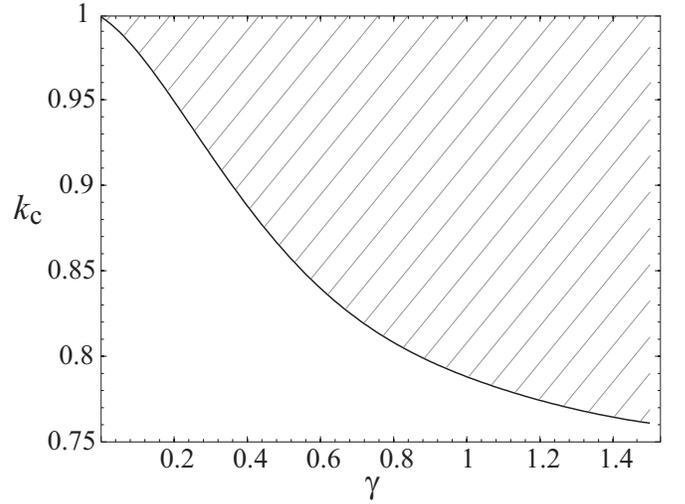}
\caption{\label{kcr}
The critical value of the elliptic modulus as a function of $\gamma$. 
Shaded is the stability domain;  for $k<k_c(\gamma)$ the wave is unstable.
}
\end{figure}

Note that when $h=\gamma$, the original and scaled damping coefficients
coincide: $\gamma={\tilde \gamma}$. Therefore, on the 
$(\gamma,h)$-plane, the stability region is given by the ray $h=\gamma$ with $\gamma> {\tilde \gamma_c}(k)$.
The inverse function $k_c(\gamma)$ [where $k_c(\gamma_c(k))=k$]
defines the boundary of the stability region on the $(\gamma,k)$-plane (Fig.\ref{kcr}).

\section{Solutions $cn^+$ and $cn^-$: $h> \gamma$}

In this section,  we calculate the deformation of the linear spectrum 
associated with the cnoidal waves $\mathrm{cn}^+$ and $\mathrm{dn}^+$ as 
$h$ grows from the value $h=\gamma$.

It is not difficult to evaluate the change of the symplectic spectrum
as $h$ deviates from $\gamma$ and hence
the parameter $\mathcal E$ deviates from zero. The corresponding spectrum
coincides with the spectrum of the undamped NLS with nonzero parametric
driving.

In addition to  expanding  ${\vec {\mathcal Y}}$ and $\lambda$ as in \eqref{T1A}, we
let ${\mathcal E}= {\mathcal E}_0 \kappa^2$ in the operator \eqref{2.7}
[which, in turn, is a matrix element of the operator \eqref{S30}].
Substituting these expansions in 
 equation \eqref{T1}, the orders $\kappa^0$ and $\kappa^1$ produce the same equations as 
  in section \ref{CN_E0_pert}, with the same set of
  solutions. The first difference from the case ${\mathcal E}=0$ arises
  at the order $\kappa^2$ where the equation \eqref{T100} is replaced with 
 \[
 \mathcal{L}_0 v_2= 2i \partial_X v_1-(1\mp {\mathcal E}_0)v_0+ \lambda_2 u_0+ \lambda_1 u_1, 
 \]
and the corresponding solvability condition \eqref{T102} with 
\[
2i\langle q | \partial_X v_1\rangle - \mathcal{B} (1\mp {\mathcal E}_0)
\langle q | q\rangle + \lambda_1 \langle q| u_1\rangle=0.
\]
[On the other hand, the solvability condition \eqref{T101} is not changed.]
As a result, the only matrix element of the matrix $\mathcal M$ that 
is altered by allowing a nonzero $\mathcal E$, is $n_{22}$ which becomes
\[
n_{22}=  \frac{4k}{\sqrt{2k^2-1}} \frac{K^2}{K_k} \pm {\mathcal E}_0N.
\]
The quartic equation \eqref{biqua} is replaced with
\[
a_0 \lambda_1^4 + (b_0 \mp {\mathcal E}_0 b_1) \lambda_1^2 +(c_0 \mp {\mathcal E}_0 c_1)=0,
 \]
where
\[
b_1= -2(LN)_k L_k,
\quad
c_1= -\frac{4L N^2}{(1-k^2) (2k^2-1)^3}
\]
are two negative constants.

Defining $\mu$ such that $\lambda_1= {\mathcal E}_0^{1/2} \mu$, the
quartic equation becomes
\be
 a_0 \mu^4 + 
 \left(\frac{b_0}{{\mathcal E}_0} \mp  b_1\right) \mu^2 +\frac{c_0}{{\mathcal E}_0^2} \mp \frac{c_1}{{\mathcal E}_0}=0.
 \label{qua} \ee
 Here ${\mathcal E}_0$ is a positive parameter. 
 
\begin{figure}[h]
 \begin{center}
 \includegraphics*[scale=1.0]{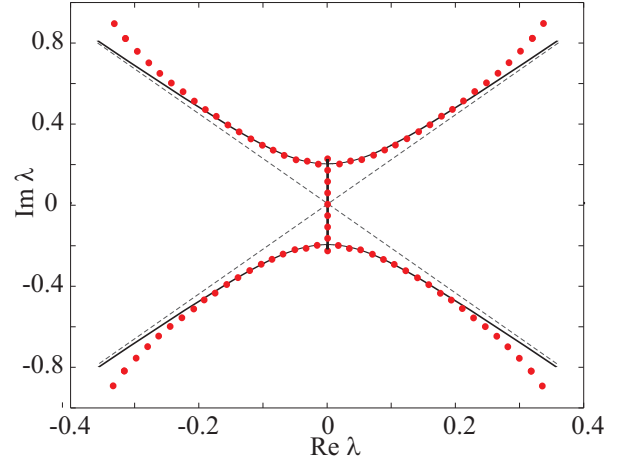}
 \caption{Solid curves: trajectories of the four roots of  Eq.\eqref{qua} 
 as ${\mathcal E}_0$ is varied from $0$ to $\infty$. The dashed lines 
 are plotted at the angles $\pm \phi$ to the horizontal, with $\phi$ as in \eqref{T11}.  Dots: numerically obtained symplectic
 eigenvalues.  In this plot,  $k^2=0.8$.  ${\mathcal E}=0.01$ }
 \label{nonzero_e}
 \end{center}
 \end{figure}

Consider the top sign in \eqref{qua}, that is, consider eigenvalues
pertaining to the cnoidal wave $cn^+$.
For ${\mathcal E}_0 = \infty$,
  two roots of equation \eqref{qua} are equal to zero,
  $\mu = 0$, while the other two roots take  opposite pure imaginary values: 
  $\mu = \pm i (-b_1/a_0)^{1/2}$. 
  As ${\mathcal E}_0$ decreases from large values, two pairs of opposite roots move
  along the imaginary axis, collide pairwise, and appear into the complex plane, forming
  a quadruplet $\pm \mu,    \pm \mu^*$.  
  As ${\mathcal E}_0 \to 0$, the four complex roots diverge from the origin along 
  the straight lines making the angles $\pm \phi$ to the horizontal, with $\phi$ as in \eqref{T11}. 
   The trajectories of the roots of the equation \eqref{qua} 
 as ${\mathcal E}_0$ is varied are shown in Fig.\ref{nonzero_e}. 
 
 The symplectic eigenvalues $\lambda$ are related to the roots $\mu$
 by the scaling $\lambda =  {\mathcal E}^{1/2} \mu$. 
 Thus for any small $\mathcal E$, the small-$\lambda$ part of the symplectic spectrum of the 
 $cn^+$ wave is obtained by scaling  down Fig.\ref{nonzero_e}. 
 
\appendix
\section{Some useful inequalities involving elliptic integrals}

{\bf 1.} First we show that the number of particles 
captured in one period of the $dn$ wave, eq.\eqref{Ndn},  is a monotonically
decreasing function of $k$. 
A quick way to establish this is to note that 
the derivative
\[
N_k= \frac{2}{k (2-k^2)^{3/2}} [2E-(2-k^2)K]
\]
can be related to the 
manifestly positive integral\[
\int_0^K \frac{\mathrm{sn}^2(\xi, k)  \mathrm{cn}^2(\xi, k)}{\mathrm{dn}^2(\xi, k)} d\xi
= -\frac{1}{k^2} [2E-(2-k^2)K].
\]

{\bf 2.} Next, we prove that the derivative
\be
\frac{d}{dk} (L_{\mathrm dn}N)=
\frac{4}{k(1-k^2)} [E^2-(1-k^2)K^2]
\label{LNk}
\ee
with $N$ as in \eqref{Ndn}, 
is positive. To this end, it is sufficient to observe (a) that the expression in the 
right-hand side of \eqref{LNk} equals zero at $k=0$, and (b) 
that the derivative of this expression is positive:
\[
\frac{d}{dk} [E^2-(1-k^2)K^2]= \frac{2}{k} (E-K)^2.
\]

{\bf 3.} 
Since $(LN)_k>0$ while $N_k<0$, one is led to conclude that $L_k>0$,
the period of the $dn$ wave is a monotonically
growing function of $k$.  An independent way to see this is to note that 
the derivative
\[
\frac{dL_{\mathrm dn}}{dk}=\frac{ 2\sqrt{2-k^2}}{k}
\left( \frac{E}{1-k^2} - \frac{2K}{2-k^2} \right)
\]
can be related to the 
manifestly positive integral
\[
\int_0^K \mathrm{sn}^2 (\xi, k) \mathrm{cn}^2 (\xi, k) d \xi=
\frac{(2-k^2) E-2(1-k^2) K}{3k^4}.
\]

{\bf 4.} Finally, we prove the inequality  
\be
\frac{k'}{k} \frac{K(k)-E(k)}{E(k)}<1
\label{A1}
\ee
for all $\frac12< k^2 <1$. 
Here $k'=\sqrt{1-k^2}$ is the complementary modulus of the elliptic integrals.

We start by writing \eqref{A1} as 
\be
g(k)<1,
\label{A2}
\ee
where we have defined
\[
g(k)= -  \frac{k'}{E(k)} \frac{dE}{dk}.
\]

At the point $k_0=1/\sqrt{2}$, the integral $K(k)$ has the property $dK/dk=K$
which allows to relate $K$ and $E$,
\[
K= \frac{1}{{k'}^2(1+k)} E,
\]
and subsequently evaluate $g(k_0)$:
\[
g(k_0)= \frac{\sqrt{2}-1}{\sqrt{2}+1}.
\]
Therefore, at the point $k_0$ the value of the function
$g(k)$ is smaller than 1 and will remain smaller 
than 1 in some neighbourhood of this point.
We will now show that this neighbourhood extends all the way to $k=1$.

Assume the contrary, that is, assume that there is a point $k_*$,
$k_0< k_*<1$,  such that 
\be
g(k_*) =1.
\label{A8} \ee
Using the hypergeometric equation satisfied by $E(k)$, 
\[ 
k {k'}^2  \frac{d^2E}{dk^2} + {k'}^2 \frac{dE}{dk} +kE=0,
\]
we obtain  a Riccati equation
\be
\frac{df}{dk}= \frac{f^2}{k} + \frac{k}{{k'}^2}
\label{A3} \ee
for  the function
\[
f(k)= \frac{k}{k'}  g(k).
\]
Integrating both sides of \eqref{A3}
from $k_0$ to $k_*$ we obtain
\be
f(k_*)-f(k_0)= \int_{k_0}^{k_*} \left[ \frac{f^2}{k} + \frac{k}{{k'}^2} \right] dk.
\label{A5} 
\ee
Since $f(k) < k/k'$ in the interval $(k_0, k_*)$, the integral admits a simple 
bound:
\be
\int_{k_0}^{k_*} \left[ \frac{f^2}{k} + \frac{k}{{k'}^2} \right] dk < \int_{k_0}^{k_*} \left[ \frac{1}{k} \left(\frac{k}{k'}\right)^2+ \frac{k}{{k'}^2} \right] dk.
\label{A4}
\ee
The integral in the right-hand side of \eqref{A4} is tabular
and hence \eqref{A5} gives
\be
f(k_*)<  1+ \ln \frac{k_0^2}{1-k_*^2},
\label{A6} \ee
where we have also used $f(k_0)<1$.

On the other hand, one can readily show that 
the function 
\[
F_1(k)=1+ \ln \frac{k_0^2}{1-k^2}
\]
is smaller than $F_2(k)=k/k'$ 
for all $k>k_0$. 
[Indeed, we have $F_1(k_0)=F_2(k_0)=1$
but $dF_1/dk< dF_2/dk$ for all $0<k<1$.] In particular, 
$F_1(k_*)$ is smaller than $F_2(k_*)$, that is, 
\be
1+ \ln \frac{k_0^2}{1-k_*^2} < \frac{k_*}{k_*'},
\label{A7} \ee
where $k_*'=\sqrt{1-k_*^2}$. Using \eqref{A7}, the inequality \eqref{A6} becomes
$f(k_*)<  k_*/k_*'$ or, equivalently,
\be
g(k_*)<1
\label{A9} \ee
which contradicts \eqref{A8}.

The contradiction proves that 
no point $k_*<1$ satisfying \eqref{A8} can exist. Therefore the 
inequality \eqref{A2} (and thus, \eqref{A1}) 
remain valid for all $k$ between $k_0$ and 1.

\section*{Acknowledgements}

M.M. was supported by the National Research Foundation (NRF) of
South Africa and the National Institute for Theoretical Physics
(NITheP).

\end{document}